\documentclass[conference]{IEEEtran}
\IEEEoverridecommandlockouts
\usepackage{cite}
\usepackage{amsmath,amssymb,amsfonts}
\usepackage{algorithmic}
\usepackage{graphicx}
\usepackage{url}
\usepackage{textcomp}
\usepackage{xcolor}
\usepackage{color}
\usepackage{tikz}
\usepackage{svg}
\usepackage{pgfplots}
\pgfplotsset{compat=1.18}
\usepackage{subfigure}
\usepackage{booktabs}
\usepackage{float}
\def\BibTeX{{\rm B\kern-.05em{\sc i\kern-.025em b}\kern-.08em
    T\kern-.1667em\lower.7ex\hbox{E}\kern-.125emX}}
\begin{document}

\title{A Mel Spectrogram Enhancement Paradigm Based on CWT in Speech Synthesis}
\author{\IEEEauthorblockN{1\textsuperscript{st} Guoqiang Hu}
\IEEEauthorblockA{\textit{International School} \\
\textit{Jinan University}\\
Guangzhou, China \\
violetpark6567@gmail.com}

\and
\IEEEauthorblockN{2\textsuperscript{nd} Huaning Tan}
\IEEEauthorblockA{\textit{International School} \\
\textit{Jinan University}\\
Guangzhou, China \\
tanhuaning24@gmail.com}

\and
\IEEEauthorblockN{3\textsuperscript{rd} Ruilai Li}
\IEEEauthorblockA{\textit{International School} \\
\textit{Jinan University}\\
Guangzhou, China \\
liruilaimz@outlook.com}
}

\maketitle

\begin{abstract}
Acoustic features play an important role in improving the quality of the synthesised speech. Currently, the Mel spectrogram is a widely employed acoustic feature in most acoustic models. However, due to the fine-grained loss caused by its Fourier transform process, the clarity of speech synthesised by Mel spectrogram is compromised in mutant signals. In order to obtain a more detailed Mel spectrogram, we propose a Mel spectrogram enhancement paradigm based on the continuous wavelet transform (CWT). This paradigm introduces an additional task: a more detailed wavelet spectrogram, which like the post-processing network takes as input the Mel spectrogram output by the decoder. We choose Tacotron2 and Fastspeech2 for experimental validation in order to test autoregressive (AR) and non-autoregressive (NAR) speech systems, respectively. The experimental results demonstrate that the speech synthesised using the model with the Mel spectrogram enhancement paradigm exhibits higher MOS, with an improvement of 0.14 and 0.09 compared to the baseline model, respectively. These findings provide some validation for the universality of the enhancement paradigm, as they demonstrate the success of the paradigm in different architectures.
\end{abstract}

\begin{IEEEkeywords}
Mel spectrogram, Speech Synthesis, Fine Grainedness, Continuous Wavelet Transform
\end{IEEEkeywords}

\section{Introduction}
Currently, Text-to-speech(TTS) is a research technology that can turn text to speech, which has a wide range of applications in daily life. These include telephone customer service, intelligent home appliances, intelligent reading, and so on. At present, the most common methods of speech synthesis include the resonance peak parameter method, the splicing method, the statistical parameter method \cite{2_klatt1987review}, and methods based on deep learning\cite{3_lecun2015deep}, among others. Nevertheless, the resonance peak parameter method of speech synthesis necessitates a substantial quantity of training data. However, it should be noted that the resonance peak parameter method of speech synthesis needs high quality dataset in order to achieve an effective synthesis effect. Furthermore, the expressive power of the method is also limited in certain instances, particularly when it comes to the nuances of speech. Furthermore, both waveform splicing and statistical parametric speech synthesis present challenges. Waveform splicing is hindered by the difficulty of achieving a seamless transition between segments, while statistical parametric speech synthesis is prone to voice breaks or noise. Additionally, both approaches require significant computational resources, exhibit limited generalisation abilities, and are slow to synthesise.

The power of AI has led to the emergence of deep neural network-based speech synthesis models, which have become a dominant force in the field due to their impressive ability to generate realistic-sounding speech and their capacity for generalisation. In order to generate the corresponding speech based on the text, it is first necessary to create a text front-end to pre-process it. This is followed by the implementation of the text to Mel Spectrogram mapping, which is carried out using the acoustic model. The quality of the generated Mel Spectrogram has a significant impact on the final speech, and this step is achieved through the use of a vocoder. The majority of speech synthesis models adhere to this fundamental process. Subsequent to this, lots of deep neural network acoustic models have been proposed, each with different features. This includes Deepvoice, Tacotron2, Transformer TTS, Fastspeech2, VITS, and so forth. These models have made a significant contribution to the development of the speech synthesis field. Furthermore, the models proposed in recent years, such as NaturalSpeech series, have abandoned the Mel spectrogram, and instead use the waveform decoder to obtain the original waveform. This indicates that the Mel spectrogram is being increasingly replaced due to issues such as low subtlety.

From the perspective of signal processing, the limitations of the Mel spectrogram can be attributed to two primary factors: the fixed resolution and the fine-grained degradation resulting from the fixed window length, as well as the difficulty in fitting mutating signals with the Fourier transform basis functions. It is also worth noting that the short-time Fourier transform responds relatively slowly to changes in the voice signal, resulting the need for longer time window to capture these changes. Consequently, the Mel spectrogram enhancement paradigm proposed herein entails the prediction of the Mel spectrogram output from the Mel spectrogram decoder, generated using the continuous wavelet transform, in addition to enhancement by Post-Net reconstruction. CWT's dynamic window size allows for more effective handling of transient variations in the speech signal, by providing different resolutions at different frequencies, thus improving the capture of time-frequency of the voice. And the additional auxiliary task \cite{11_zhang2021survey} will encourage the Mel spectrogram to learn in a more detailed manner, which will subsequently enhance the expressiveness and clarity of the speech synthesised.

This paper comprises four sections. The first section examines several acoustic models that rely on the Mel spectrogram and the application of the CWT in Fastspeech2. Second section presents the theoretical properties of the continuous wavelet transform, demonstrating the theoretical feasibility of this enhancement paradigm. The third section introduces the specific components and detailed processes of the Mel spectrogram enhancement paradigm, including the Mel spectrogram decoder, the post-processing network and the CWT-Net. Part IV presents the model architectures of Tacotron2 and Fastspeech2, which have been enhanced using the Mel spectrogram paradigm. It also outlines the data processing that occurs during the experimental process and presents the experiment result. In the fifth section, the advantages and disadvantages of the enhanced models are analysed and discussed, as well as the future development trend of acoustic models, which has been introduced in this paper.

\section{Related Work}
\begin{figure*}[t!]
\caption{Mel Spectrogram Enhancement Paradigm Framework}
\centering
\includegraphics[width=0.6\linewidth]{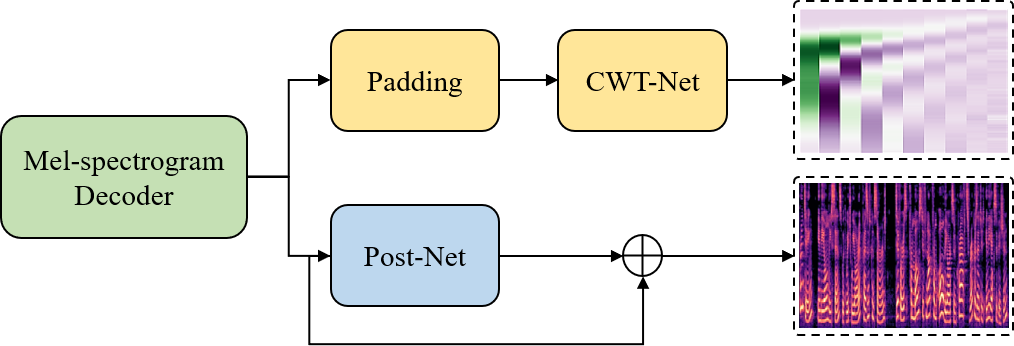}
\label{fig1}
\end{figure*}
\subsection{Text-to-Speech}
Currently, the majority of research in the field of speech synthesis employs deep learning approaches. The advent of diverse network architectures, including CNN and LSTM, has greatly facilitated the extraction of deep speech features. This has led to the emergence of numerous speech synthesis systems that employ these combinations to construct acoustic models. For instance, Baidu's Deepvoice series of codecs are constructed using CNNs, while Tacotron2 employs a variant of RNN to autoregressively generate a frame-by-frame Mel spectrogram. Furthermore, Transformer, the most powerful model currently available, also plays an important role in speech synthesis. Finally, some speech synthesis began to explore alternative solutions beyond traditional signal processing. VALL-E\cite{30_wang2023neural} represents a prime example of this shift, treating speech synthesis as a conditional language modelling task by converting speech into discrete codes. In addition to its use in speech synthesis, Mel spectrogram is employed in the field of speech translation\cite{12_le2024comsl}.

\subsection{Acoustic Model}
The acquisition of acoustic features as intermediary features can be accomplished through the implementation of a multitude of signal processing techniques, which play a pivotal role within the domain of TTS technologies. In the field of signal processing, wavelet transform\cite{14_rioul1992fast} offers a valuable tool for the analysis and processing of complex signals. Its fine analysis and numerous wavelet functions provide a comprehensive approach to understanding and manipulating these signals. Furthermore, the Fourier transform\cite{13_griffin1984signal} is frequently employed in the processing of speech signals with the objective of generating Mel spectrogram that more accurately reflect the perceptual properties of the human ear. Nevertheless, in recent years, research has been conducted with the goal to improve the integration of speech into large language models. This has involved the conversion of speech into discrete codes\cite{15_shen2024acoustic}. Among these, Encodec\cite{16_defossez2022high}, a pre-trained audio compression model, is frequently employed as a tokenizer, enabling the direct decoding of speech from discrete codes generated by a quantization layer, thereby achieving speech quality beyond that of the Mel spectrogram\cite{17_wang2023neural}. Consequently, there is a pressing necessity for the development of a paradigm that will enhance the Mel spectrogram, in order to facilitate the advancement of speech synthesis in the context of the current era.

\subsection{Mel Spectrogram}
Numerous studies have been conducted on the Mel spectrogram, which is used in speech representation and TTS. However, the majority of these studies are conducted at the data level. Yeongtae Hwang\cite{18_hwang2020mel} employs frequency warping, loudness control and time length regulation on the Mel spectrogram to obtain diverse data. Hao Meng\cite{19_meng2019speech} achieves superior results in the transformation of emotional dialogues by extracting 3D log Mel Spectrogram. All of the aforementioned acoustic models utilise Mel spectrogram as intermediate features. However, the qualities of the continuous wavelet transform do not appear to be particularly advantageous in the field of speech synthesis. Nevertheless, the initial utilisation of the wavelet transform to extract fundamental frequencies in Fastspeech2 exemplifies the advantages of the wavelet transform \cite{20_ribeiro2016wavelet} in capturing subtle changes.

\subsection{Tacotron2 and Fastspeech2}
The fundamental component of Tacotron2 is the mapping of text into Mel spectrogram. For each input, the encoder first maps each phoneme into a 512-D vector using a 512-D embedding, which is then passed through several 1-D CNNs. Output of the CNN is then fed into a single-layer Bi-LSTM. The attention mechanism, which is part of the Tacotron2, employs a position-sensitive attention mechanism that requires both its own attention vectors from the previous time step and output of decoder from previous time step. Output of the previous time step decoder is initially processed by a pre-net comprising two fully connected layers, after which it is fed into the attention module. The context vectors computed by the attention module are input to the decoder LSTM, which together with the hidden state of the decoder predicts the Mel spectrogram. Finally, Mel spectrogram generated frame-by-frame and spliced is passed through 5 layers CNN in Post-Net to enhance the overall reconstruction. Conversely, since Tacotron2 is an autoregressive architecture, the decoder hidden state is also fed into another fully-connected layer that outputs a gated stopping probability. This is used to determine whether the decoder should cease operation at the current step.

The FastSpeech2 model adheres to the general framework of FastSpeech, comprising an encoder, a variable information adapter and a vocoder. The variable information adapter is employed to extract variable information from speech, including duration, pitch, volume, and so forth, with the goal to address the problem that the same text can be pronounced in a multitude of styles. The variable information adapter comprises a variety of variable information predictors, including duration predictor, pitch predictor and energy predictor. These predictors correspond to different speech feature information. In the training stage, the model inputs the encoded character vectors into the encoder, together with the speech features as inputs to the decoder, and trains each predictor concurrently. In the model's inference stage, the variable information predictors are employed to predict the corresponding speech features. Similarly to Tacotron2, Fastspeech2 proposes a Mel spectrogram Enhancement Paradigm, which is inspired by the Mel spectrogram Decoder. This is because the Mel spectrogram generated by the Mel spectrogram Decoder is reconstructed and enhanced by Post-Net during the actual training of the model.

\section{Analysis of CWT and Fourier Transform}
Fourier transform and wavelet transform\cite{26_hess1996wavelets} are two widely used transforms in traditional signal processing. The former is often employed in traditional fields to extract acoustic features of speech. However, the Fourier transform is not sufficiently sensitive to the temporal information of a signal to provide sufficiently fine-grained local information. In particular, the properties of the Fourier transform's windowing operation and basis functions present difficulties in fitting mutant signals.

One of the variants of the wavelet transform, CWT\cite{21_mallat1992singularity}, can be adapted to different frequency domain parts of the voice signal by adjusting scale of wavelet function. The resulting time-frequency localised features can be employed as an efficient representation of the data. Moreover, it enables the signal to be captured simultaneously in time-frequency domains, which is beneficial for non-stationary signals such as bursts or transients. The basic formula for performing the CWT on a voice signal $f(t)$ is given in Eq. (\ref{Eq:1}):

\begin{footnotesize} 
\begin{equation}
W(x,y)=\int_{-\infty }^{+\infty } f(t)\exp(-i\omega _{0} \frac{(t-y)}{x} )\exp(-\frac{(t-y)^{2} }{2x^{2} })dt
\label{Eq:1}
\end{equation}
\end{footnotesize}

\noindent In this equation, $a$ denotes the scale ratio, $b$ indicates the time delay factor, and $f(t)$ stands for the voice signal.
\vspace{12pt}

It should be noted that there are several options for wavelet basis functions, the most commonly used being the morelet wavelet\cite{25_zhou2023application}, whose formula is shown in Eq. (\ref{Eq:2}). In practical processing, we should consider the characteristics of the signal to be analysed as a way to choose the most appropriate wavelet basis function.

\begin{equation}
\psi (t)=\exp(i\omega _{0}t)\exp(-\frac{t^{2} }{2}  )  
\label{Eq:2}
\end{equation}

\section{Proposed Approach}
The framework of our proposed Mel spectrogram Enhancement Paradigm is depicted in Fig. \ref{fig1}, which comprises three components: The Mel spectrogram Decoder, CWT-Net and Post-Net are the three components of the proposed Mel spectrogram Enhancement Paradigm. Subsequently, the Mel spectrogram obtained from the Mel spectrogram Decoder is passed through the CWT-Net and Post-Net, respectively. The former is designed to predict finer wavelet spectrograms, thereby forcing the Mel Spectrograms to transform in a more delicate direction \cite{22_sener2018multi}. The latter draws on the majority of speech synthesis strategies to reconstruct and enhance the previous Mel spectrogram. The loss of the wavelet spectrogram is calculated, and the overall loss of the model applying our proposed method is shown in Equations (\ref{Eq:3}) and (\ref{Eq:4}), respectively. From the perspective of multi-task learning, our proposed method incorporates the wavelet spectrogram as an additional task to reconstruct the loss function of the speech synthesis model. This will facilitate the convergence of the model parameters towards finer grains. In the following section, we will present the specific composition of CWT-Net.

\begin{equation}
    Loss_{Wavelet}=\frac{1}{n} \sum_{i=1}^{n} (\hat{x} -x)^{2}
\label{Eq:3}
\end{equation}
\begin{equation}
    Loss_{Total}  = Loss_{Baseline} + Loss_{Wavelet}
\label{Eq:4}
\end{equation}

\subsection{CWT-Net}
The CWT-Net comprises a number of one-dimensional convolutional neural networks and linear layers. The input is a Mel Spectrogram generated by a Mel Spectrogram decoder. The first dimension of the Mel spectrogram is related to Mel filters\cite{31_imai1983mel}, which typically ranges from 80 to 128. The second dimension is proportional to the duration of voice signal. In order to maintain the dimensions of the input to the model constant, a padding\cite{27_nguyen2019distribution} operation is performed on the Mel spectrograms of all the data. Subsequently, a one-dimensional convolutional neural network and linear layers are employed to progressively capture high-level, fine-grained features from the global Mel spectrogram. In addition, multiple normalisation layers and activation functions are employed to mitigate the effects of gradient vanishing and facilitate loss convergence.

\section{Experiments and Results}

\begin{figure}[t!]
\caption{Tacotron2 using Mel Spectrogram Enhancement Paradigm}
\centering
\includegraphics[width=0.6\linewidth]{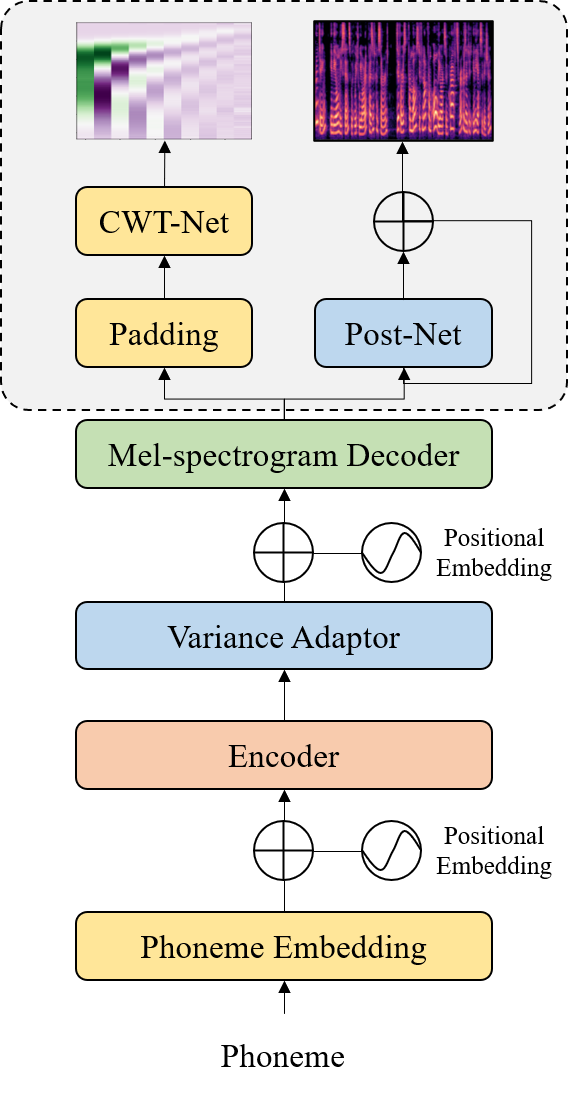}
\label{fig2}
\end{figure}

\begin{figure}[t!]
\caption{Fastspeech2 using Mel Spectrogram Enhancement Paradigm}
\centering
\includegraphics[width=\linewidth]{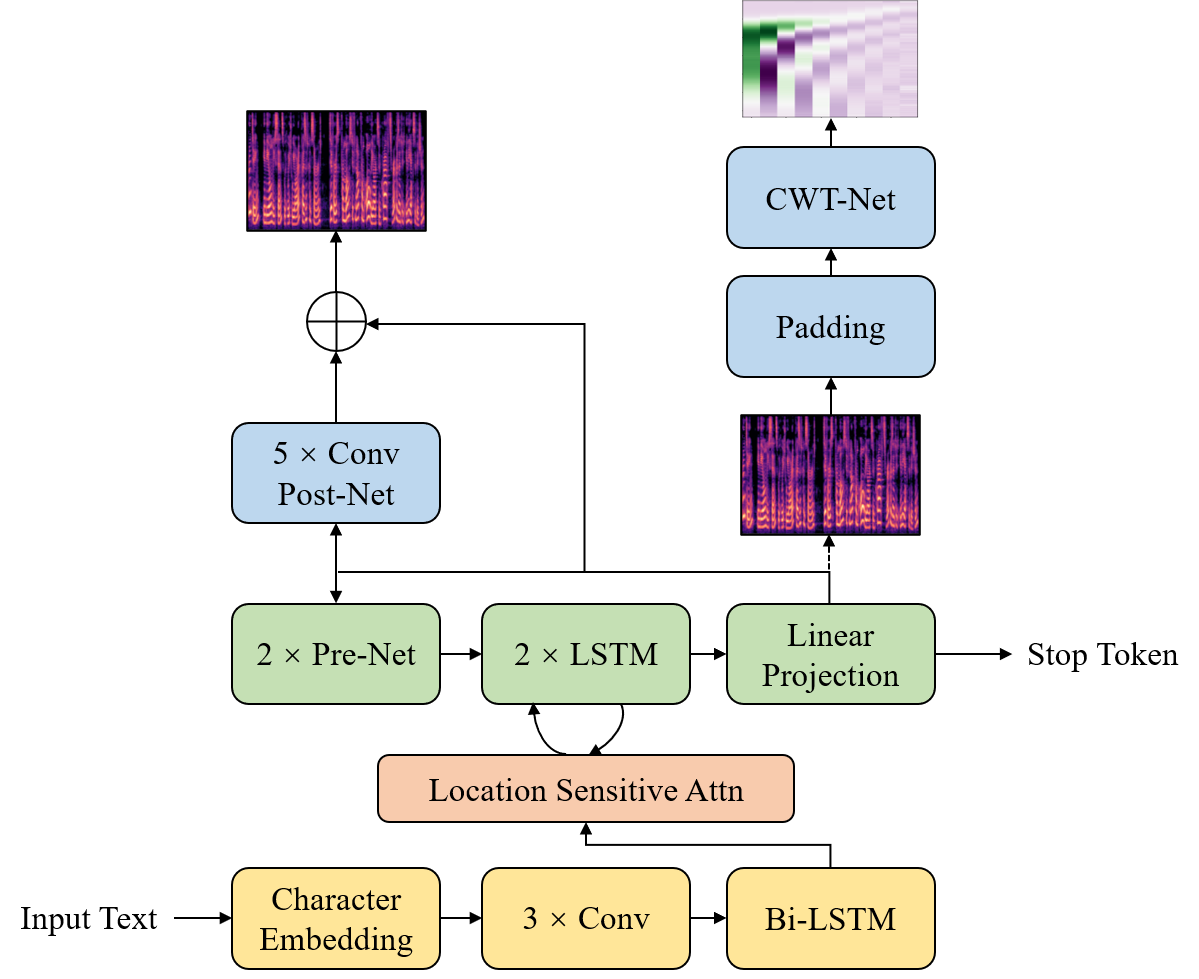}
\label{fig3}
\end{figure}
\subsection{Dataset and Preprocessing}
In the course of the experiments, the model was validated using its original English dataset, LJspeech\cite{23_ljspeech17}, which comprises 13100 sentences with duration of approximately twenty four hours. In the preprocessing phase, the convention of feature extraction was followed for each speech file in order to obtain its Mel spectrogram. Furthermore, signal processing tools, such as the continuous wavelet transform of the speech segments, are required to obtain more detailed wavelet spectrogram. Due to the high-dimensional\cite{28_donoho2000high} nature of the wavelet spectrogram, which makes the training process difficult to converge, we employ truncated singular value decomposition\cite{24_golub1971singular} to downscale it. The downscaled wavelet spectrogram is then used as the prediction object.

\subsection{Result Analysis}

In order to demonstrate the universality of our proposed Mel spectrogram enhancement paradigm, we selected Tacotron2 and Fastspeech2 as the baseline models in AR-TTS and NAR-TTS, respectively, for our experiments. The framework structures of these models, as they were applied with our proposed approach, are shown in Figures \ref{fig2} and \ref{fig3}. The objective of this study is to evaluate the speech synthesised. To ensure consistency and focus on the audio quality evaluation, the same text input was provided for each model. A total of 30 scorers, comprising foreign students, Hong Kong students and mainland students, participated in this experiment. Each scorer was asked to provide a rating ranging from 0 to 5 for each audio.

Following the removal of certain maxima and minima, the result of the experiment are presented in Table 1. It can be observed that the speech synthesis system utilising the Mel spectrogram Enhancement Paradigm receives significantly higher ratings from the testers, as evidenced by the higher mean opinion score (MOS) \cite{29_streijl2016mean}. This indicates that the proposed Mel spectrogram Enhancement Paradigm is effective in enhancing the fine-grainedness of the obtained Mel spectrogram, both in AR-TTS and NAR-TTS, which subsequently improves the clarity of the synthesised speech.
\begin{table}[H]
\caption{Comparison of the Evaluated MOS for Baseline and Enhancement Paradigms}
\centering
\begin{tabular}{cc}
\toprule 
Model & Mos \\  
\midrule 
Tacotron2 & 3.70 \\  
Tacotron2+Enhancement & 3.84 \\  
Fastspeech2 & 3.83 \\  
Fastspeech2+Enhancement & 3.92 \\  
\bottomrule 
\end{tabular}
\end{table}

\section{Summary}
This paper proposes a Mel spectrogram enhancement paradigm that can effectively improve the fine-grainedness of the obtained Mel spectrogram. The enhancement paradigm can be applied to any TTS synthesis system containing a Mel spectrogram decoder and a post-processing network. Specifically, we force the Mel spectrogram reconstructed by its post-processing network to change in a more fine-grained direction by extracting a more fine-grained wavelet spectrogram from the Mel spectrogram synthesised by the Mel spectrogram decoder. To demonstrate the universality of this paradigm, we selected Tacotron2 and Fastspeech2 for experiments at AR-TTS and NAR-TTS, respectively, and succeeded. Nevertheless, with the advancement of technology, some TTS have ceased to utilise Mel spectrogram for the synthesis of speech, such as VALL-E. Consequently, the potential for enhancing other features, such as the encoder, in accordance with the principles outlined in this paper will be the subject of our future research.


\bibliographystyle{ieeetr}
\bibliography{Ref.bib}

\begin{thebibliography}{10}

\bibitem{2_klatt1987review}
D.~H. Klatt, ``Review of text-to-speech conversion for english,'' {\em The Journal of the Acoustical Society of America}, vol.~82, no.~3, pp.~737--793, 1987.

\bibitem{3_lecun2015deep}
Y.~LeCun, Y.~Bengio, and G.~Hinton, ``Deep learning,'' {\em nature}, vol.~521, no.~7553, pp.~436--444, 2015.

\bibitem{11_zhang2021survey}
Y.~Zhang and Q.~Yang, ``A survey on multi-task learning,'' {\em IEEE Transactions on Knowledge and Data Engineering}, vol.~34, no.~12, pp.~5586--5609, 2021.

\bibitem{30_wang2023neural}
C.~Wang, S.~Chen, Y.~Wu, Z.~Zhang, L.~Zhou, S.~Liu, Z.~Chen, Y.~Liu, H.~Wang, J.~Li, {\em et~al.}, ``Neural codec language models are zero-shot text to speech synthesizers,'' {\em arXiv preprint arXiv:2301.02111}, 2023.

\bibitem{12_le2024comsl}
C.~Le, Y.~Qian, L.~Zhou, S.~Liu, Y.~Qian, M.~Zeng, and X.~Huang, ``Comsl: A composite speech-language model for end-to-end speech-to-text translation,'' {\em Advances in Neural Information Processing Systems}, vol.~36, 2024.

\bibitem{14_rioul1992fast}
O.~Rioul and P.~Duhamel, ``Fast algorithms for discrete and continuous wavelet transforms,'' {\em IEEE transactions on information theory}, vol.~38, no.~2, pp.~569--586, 1992.

\bibitem{13_griffin1984signal}
D.~Griffin and J.~Lim, ``Signal estimation from modified short-time fourier transform,'' {\em IEEE Transactions on acoustics, speech, and signal processing}, vol.~32, no.~2, pp.~236--243, 1984.

\bibitem{15_shen2024acoustic}
F.~Shen, Y.~Guo, C.~Du, X.~Chen, and K.~Yu, ``Acoustic bpe for speech generation with discrete tokens,'' in {\em ICASSP 2024-2024 IEEE International Conference on Acoustics, Speech and Signal Processing (ICASSP)}, pp.~11746--11750, IEEE, 2024.

\bibitem{16_defossez2022high}
A.~D{\'e}fossez, J.~Copet, G.~Synnaeve, and Y.~Adi, ``High fidelity neural audio compression,'' {\em arXiv preprint arXiv:2210.13438}, 2022.

\bibitem{17_wang2023neural}
C.~Wang, S.~Chen, Y.~Wu, Z.~Zhang, L.~Zhou, S.~Liu, Z.~Chen, Y.~Liu, H.~Wang, J.~Li, {\em et~al.}, ``Neural codec language models are zero-shot text to speech synthesizers,'' {\em arXiv preprint arXiv:2301.02111}, 2023.

\bibitem{18_hwang2020mel}
Y.~Hwang, H.~Cho, H.~Yang, D.-O. Won, I.~Oh, and S.-W. Lee, ``Mel-spectrogram augmentation for sequence to sequence voice conversion,'' {\em arXiv preprint arXiv:2001.01401}, 2020.

\bibitem{19_meng2019speech}
H.~Meng, T.~Yan, F.~Yuan, and H.~Wei, ``Speech emotion recognition from 3d log-mel spectrograms with deep learning network,'' {\em IEEE access}, vol.~7, pp.~125868--125881, 2019.

\bibitem{20_ribeiro2016wavelet}
M.~S. Ribeiro, O.~Watts, J.~Yamagishi, and R.~A. Clark, ``Wavelet-based decomposition of f0 as a secondary task for dnn-based speech synthesis with multi-task learning,'' in {\em 2016 IEEE International Conference on Acoustics, Speech and Signal Processing (ICASSP)}, pp.~5525--5529, IEEE, 2016.

\bibitem{26_hess1996wavelets}
N.~Hess-Nielsen and M.~V. Wickerhauser, ``Wavelets and time-frequency analysis,'' {\em Proceedings of the IEEE}, vol.~84, no.~4, pp.~523--540, 1996.

\bibitem{21_mallat1992singularity}
S.~Mallat and W.~L. Hwang, ``Singularity detection and processing with wavelets,'' {\em IEEE transactions on information theory}, vol.~38, no.~2, pp.~617--643, 1992.

\bibitem{25_zhou2023application}
J.~Zhou, Z.~Li, and J.~Chen, ``Application of two dimensional morlet wavelet transform in damage detection for composite laminates,'' {\em Composite Structures}, vol.~318, p.~117091, 2023.

\bibitem{22_sener2018multi}
O.~Sener and V.~Koltun, ``Multi-task learning as multi-objective optimization,'' {\em Advances in neural information processing systems}, vol.~31, 2018.

\bibitem{31_imai1983mel}
S.~Imai, K.~Sumita, and C.~Furuichi, ``Mel log spectrum approximation (mlsa) filter for speech synthesis,'' {\em Electronics and Communications in Japan (Part I: Communications)}, vol.~66, no.~2, pp.~10--18, 1983.

\bibitem{27_nguyen2019distribution}
A.-D. Nguyen, S.~Choi, W.~Kim, S.~Ahn, J.~Kim, and S.~Lee, ``Distribution padding in convolutional neural networks,'' in {\em 2019 IEEE International Conference on Image Processing (ICIP)}, pp.~4275--4279, IEEE, 2019.

\bibitem{23_ljspeech17}
K.~Ito and L.~Johnson, ``The lj speech dataset.'' \url{https://keithito.com/LJ-Speech-Dataset/}, 2017.

\bibitem{28_donoho2000high}
D.~L. Donoho {\em et~al.}, ``High-dimensional data analysis: The curses and blessings of dimensionality,'' {\em AMS math challenges lecture}, vol.~1, no.~2000, p.~32, 2000.

\bibitem{24_golub1971singular}
G.~H. Golub and C.~Reinsch, ``Singular value decomposition and least squares solutions,'' in {\em Handbook for Automatic Computation: Volume II: Linear Algebra}, pp.~134--151, Springer, 1971.

\bibitem{29_streijl2016mean}
R.~C. Streijl, S.~Winkler, and D.~S. Hands, ``Mean opinion score (mos) revisited: methods and applications, limitations and alternatives,'' {\em Multimedia Systems}, vol.~22, no.~2, pp.~213--227, 2016.

\end{thebibliography}

\vspace{12pt}

\end{document}